# Network Time Synchronization of the Readout Electronics for a New Radioactive Gas Detection System

Wolfgang Hennig, *Member, IEEE*, Vincent Thomas, Shawn Hoover and Olivier Delaune

*Abstract*— In systems with multiple radiation detectors, time synchronization of the data collected from different detectors is essential to reconstruct multi-detector events such as scattering and coincidences. In cases where the number of detectors exceeds the readout channels in a single data acquisition electronics module, multiple modules have to be synchronized, which is traditionally accomplished by distributing clocks and triggers via dedicated connections.

To eliminate this added cabling complexity in the case of a new radioactive gas detection system prototype under development at the French Atomic Energy Commission, we implemented time synchronization between multiple XIA Pixie-Net detector readout modules through the existing Ethernet network, based on the IEEE 1588 precision time protocol. The detector system is dedicated to the measurement of radioactive gases at low activity and consists of eight large silicon pixels and two NaI(Tl) detectors, instrumented with a total of three 4-channel Pixie-Net modules. Detecting NaI(Tl)/silicon coincidences will make it possible to identify each radioisotope present in the sample. To allow these identifications at low activities, the Pixie-Net modules must be synchronized to a precision well below the targeted coincidence window of 500-1000 ns. Being equipped with an Ethernet PHY compatible with IEEE 1588 and synchronous Ethernet that outputs a locally generated but system-wide synchronized clock, the Pixie-Net can operate its analog to digital converters and digital processing circuitry with that clock and match time stamps for captured data across the three modules. Depending on the network configuration and synchronization method, the implementation is capable to achieve timing precisions between 300 ns and 200 ps.

## I. Introduction

TRADITIONALLY, time synchronization between multiple channels of digital data acquisition electronics for radiation detectors is accomplished by sharing clocks, clock reset signals, and triggers through dedicated cabling, which can become quite complex [1], [2], [3], [4], [5]. For coincident events, such as the 1.17 MeV and 1.33 MeV gammas from a $^{60}$Co source detected simultaneously in separate detectors, the variation in measured time-of-arrival difference ΔT (i.e., the time resolution) can be a few hundred picoseconds full width at half maximum (FWHM) for digitization rates of 100-500 MSPS [6]. The time resolution can approach less than 10 ps FWHM for idealized signals from a pulser [6], [7] even when digitizing at less than 1 GSPS [8].

As the detector readout electronics are operated by computers linked over standard data networks, an alternative to dedicated clock distribution trees is the synchronization of clocks over the network. A current standard in network time synchronization is the IEEE 1588 precision time protocol [9] (PTP). It has been implemented on several Ethernet controllers and PHY devices (e.g. Texas Instrument's DP83640 [10]), Xilinx' Zynq processor [11], and many other devices, including commercial network switches. Precisions are reported to reach the low nanosecond range, depending on the implementation. While this precision is worse than the best reported detector time resolutions, it may still be sufficient for a range of nuclear physics applications.

However, techniques like PTP are primarily designed to synchronize clocks for processors, not for real time processing in field programmable gate arrays (FPGA) or application specific integrated circuits (ASIC). Clocks are synchronized to the nanosecond level in internal counters, but the processor can access these counters only with software limited latencies. In addition, detector waveform data, usually digitized by analog to digital converters (ADC) and captured by FPGA firmware, are not available to the processor in real time. Even if a processor could respond "immediately" to one data event, a second event can follow closer than the readout time (especially when one processor serves multiple channels).

The challenge for the use of such network synchronization techniques in detector readout electronics is therefore to integrate the synchronization with the ADC data capture and the processing of digitized detector signals in the FPGA. We report here how this has been implemented for the XIA Pixie-Net [12] and applied to a new multi-channel detector system for radioactive gases currently under development at the French Atomic Energy Commission (CEA).

The prototype detection system will be used to measure radioactive noble gases as part of environmental monitoring, such as radioactive xenon isotopes that can be released in large quantities a) during a nuclear incident, like in the Fukushima

Manuscript received June 18, 2018. This work was supported in part by the U.S. Department of Energy under Grant No. DE-SC0017223.

W. Hennig (whennig@xia.com) and S. Hoover are with XIA LLC, Hayward, CA 94544 USA. V. Thomas and O. Delaune are with CEA, DAM, DIF, F-91297.

Dai-Ichi nuclear power plant accident in 2011 [13], [14], or b) by medical isotope laboratories and facilities [15]. Four xenon radioisotopes are of interest and emit electrons and photons within a few nanoseconds [16] or less, with electron energies ranging from 0 to 915 keV and photons ranging from 30 keV X-rays to 250 keV gammas, see Table 1. As samples typically have very low activities, several radioxenon detector systems are making use of beta/gamma coincidence counting with scintillators to reduce background [17], [18], [19], [20]. More recently, silicon detectors have been used as the electron detector [21], [22] or for electron/X-ray coincidence counting [23] in order to improve the energy resolution of conversion electron (CE) peaks. The development of an ultra-compact detection system brings several constraints:
- Optimum operation at room temperature (to avoid space requirements due to the cooling system);
- Limited shielding;
- Low energy consumption system.

The choice was therefore made for a detection system using NaI(Tl) scintillators as photon detectors and a multi-channel Si detector to detect electrons and part of X-rays. Still in order to optimize its compactness, the obvious choice is a digital acquisition chain. Finally, network time synchronization of the readout electronics for each module avoids the clutter caused by a shared external clock (e.g. pulse generator) and induced wiring.

TABLE I
MAJOR EMISSION ENERGY OF THE FOUR RELEVANT RADIOXENON ISOTOPES. THE EMISSION PROBABILITIES ARE SPECIFIED IN BRACKETS. THE VALUES FOR $^{131m}$, $^{133m}$, $^{133}$XE ARE RECOMMENDED DATA TABULATED BY THE DDEP [24]. THE VALUES OF $^{135}$XE ARE THOSE RECOMMENDED BY THE NATIONAL NUCLEAR DATA CENTER [16].

| Radio-nuclide | $^{131m}$Xe | $^{133m}$Xe | $^{133}$Xe | $^{135}$Xe |
|---|---|---|---|---|
| γ-ray (keV) | 163.9 (1.94%) | 233.2 (10.1%) | 81.0 (37.0%) | 249.8 (90%) |
| K X-ray (keV) | 30.4 (54%) | 30.4 (55.9%) | 31.7 (47.6%) | 30.9 (4.9%) |
| β endpoint energy (keV) | | | 346.4 (99.1%) | 915 (96%) |
| CE (keV) | 129.4 (61.4%) | 198.7 (62.9%) | 45 (52.9%) | 213.8 (5.6%) |

## II. EXPERIMENTAL SETUP

### A. Detector System

The detector prototype consists of a gas cell surrounded by two large silicon wafers, coupled with two square NaI(Tl) detectors (Fig. 1). The gas cell has a sample volume of 30 cm³ and the silicon wafers are 500 µm thick with an active surface area of 3600 mm². In order to minimize leakage currents and thus optimize the energy resolution of the silicon detectors, each wafer has been segmented into four silicon pixels (30 x 30 mm²). This module is sandwiched between two low background NaI(Tl) detectors (83 x 263 mm² height including photomultiplier tube, 70 x 70 x 40 mm³ crystal), manufactured by Scionix Holland B.V. [25]. Each NaI(Tl) crystal is encapsulated into a low background Cu housing. The 500 µm thick entrance window is made of ultra-low background aluminum; the low-medium density (~ 2.7 g cm⁻³) of this material prevents significant absorption of X-rays emitted by the sample. The crystal volume has been optimized to effectively stop gamma rays with energy below 300 keV, while the silicon wafers' thickness is sufficient to stop any electron with a kinetic energy less than 400 keV. Geant4-based Monte Carlo simulations [26] [27] showed that this silicon thickness absorbs approximately 18 to 20% of the 30 keV photons, and does not significantly affect photons with energy higher than 80 keV (less than 1% absorbed); allowing them to pass through the silicon and to be stopped in one of the NaI(Tl) crystals.

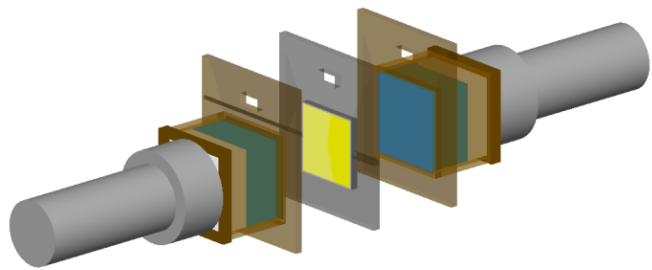

Fig. 1. Sketch of detector (Geant-4 exploded view).

In order to detect low activities (1-10 mBq/m³) of radioxenon isotopes, the coincidence measurement technique will be used to drastically reduce environmental background (which masks such activities). Detecting at least two particles within a short coincidence window allows to tag the emitting radionuclide and to distinguish it from non-coincident random background. The width of this coincidence window depends a) on the charge time collection in the detectors (which is approximately 1 µs in this case), and b) on the precision of the time synchronization of the multiple readout channels. In general, the shorter the coincidence window, the more environmental background is rejected, and the more reliable the measurement will be. However, shortening the window to less than the average time separation of background events has diminishing returns, and shortening to less than the intrinsic time resolution of the detector will lead to loss of true coincidences. In this detector system, we expect to use a window in the range of 0.5 to 1.0 µs, and require the readout channels to be synchronized with a precision of several hundred nanoseconds.

In this setup (Fig.2), two 4-channel Pixie-Net modules read out the silicon signals (one channel per pixel). An additional Pixie-Net module reads out the two NaI(Tl) signals. Data is recorded in list mode, recording time stamps, pulse height, and optionally short detector waveforms. After an acquisition, a post-processing analysis identifies if two or more particles are detected within the same coincidence window, and adds the event to a point in a 2D histogram according to the deposited energies. Events from particular isotopes thus fall into characteristic regions of interest (ROI), see Fig. 3.

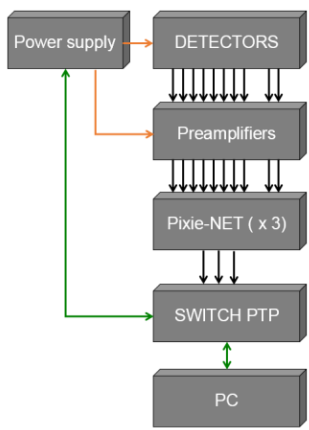

Fig. 2. Simplified diagram of detector prototype system.

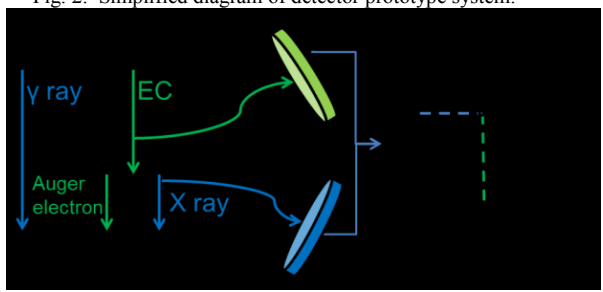

Fig. 3. Simplified $^{131m}$Xe decay scheme and resulting 2D histogram.

For a 1D histogram (direct spectrometry), the ROI of a radionuclide is an interval around the emission peak of this radionuclide. The width of the ROIs therefore depends exclusively on the energy resolution of the detector used. For a 2D histogram (coincidence spectrometry), the ROI is therefore an area whose dimensions depend on the energy resolutions of the different detectors used. In our study, by optimally setting the digital filters of the Pixie-Net, we were able to obtain an energy resolution of 6.2 and 6.3 % @ 662 keV respectively for the two NaI(Tl) detectors.

### B. Pixie-Net Readout Electronics

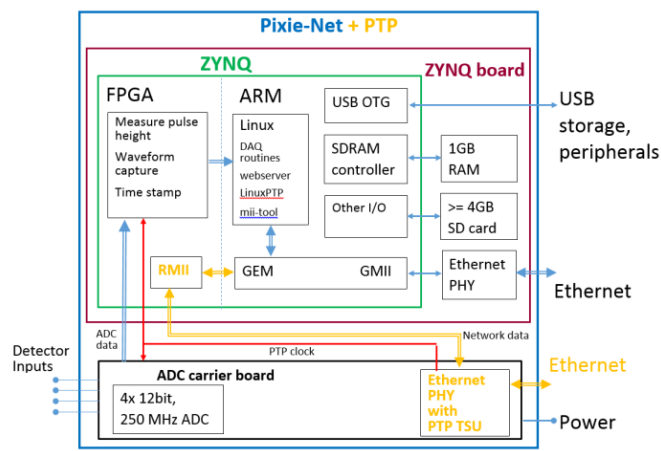

Fig. 4. Pixie-Net block diagram.

The Pixie-Net block diagram is shown in Fig. 4. It uses a Xilinx Zynq system-on-chip, which combines an ARM processor (PS) running Linux (Ubuntu 15), with an FPGA fabric (PL) processing detector pulses. The Zynq and a number of peripherals are implemented on a commercially available MicroZed board [28], and a second board implements four 250 MSPS, 12 bit ADCs that are connected to the Zynq FPGA fabric. Each Pixie-Net thus digitizes four detector signals, processes the digital data streams in the FPGA, and runs Linux programs on the Zynq's ARM processor to manage the data acquisition and communicate over the network.

To implement the PTP functionality, the Zynq Ethernet interface is connected to the DP83640 Ethernet PHY [9] which has built-in PTP hardware timestamping functions. The PHY also can be operated in synchronous Ethernet (SyncE) mode, where the clock embedded in the upstream Ethernet connection is used for local clocking. (The Zynq's built-in PTP functions make no outputs available to the FPGA and therefore are not used in this work.) The DP83640 outputs a reference clock signal synchronized to the SyncE or PTP adjusted local clock, which is connected to the FPGA and used to clock the ADCs and the FPGA pulse processing. In this manner, digitization of the detector signals is synchronized to the network PTP master clock, and all participating Pixie-Net ADCs in the entire network run on the same clock.

### C. Timing Measurements

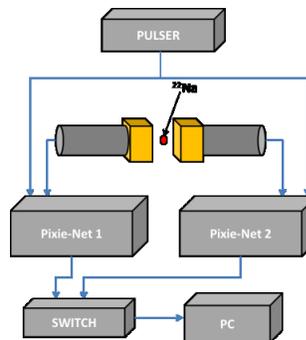

Fig. 5. Simplified diagram of the timing measurement setup.

For initial characterization of the timing performance, timing measurements were performed with two Pixie-Net modules synchronized over the network, connected to either a) two LaBr$_3$ detectors, b) an Agilent 33220A pulse generator split with identical cables, or c) the two NaI(Tl) belonging to the prototype connected to channel 0 and an Agilent 33210A pulse generator split with identical cables connected to channel 1 on each module, as shown in Fig 5. (To date, the silicon wafers are still being manufactured). The time difference ΔT between coincident gamma rays from a $^{22}$Na source or between coincident pulser signals was determined using the recorded time stamps or, when approaching the 8ns precision of the time stamps, using a constant fraction algorithm applied to captured detector waveforms [6]. In all characterization measurements, the Pixie-Net modules were configured for synchronization by PTP, SyncE, or both. The network connection was either *non-PTP* (connecting through a non-PTP switch), or *all-PTP* in which every node runs on a synchronized clock (connecting through a PTP enabled switch or connecting the two Pixie-Net modules back to back). The network switches used are listed in ref [A] – [I]. For comparison, the Pixie-Net modules were also operated with a shared clock, equivalent to the traditional dedicated clock cabling method.

### D. Simulations

The setup shown in Fig. 5 has been simulated with the Geant-4 toolkit for the NaI(Tl) detectors in order to obtain a control coincidence spectrum (Fig. 6) that can be compared to the spectra obtained with the different switches. The $^{22}$Na main coincidence ROI is framed in dashed red (inset) and corresponds to the coincidence of two 511 keV $\gamma^{\pm}$-rays, each gamma having been detected in a separate NaI(Tl) detector. Areas of secondary interest, such as Compton scattering (anti-diagonal), $\gamma^{\pm}/\!/\gamma_{(1,0)}$(Ne) (511 keV and 1274.58 keV kinetic energy, respectively) coincidences are also visible on this simulated 2D histogram.

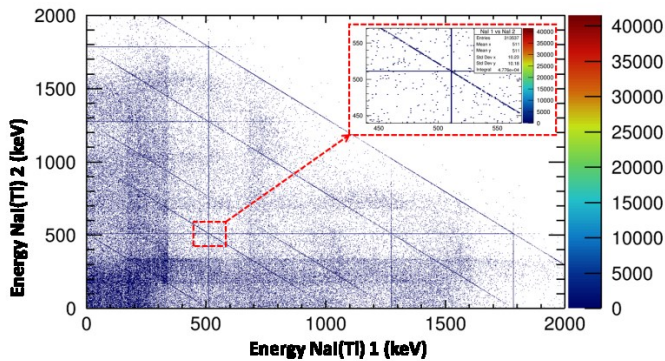

Fig. 6. $^{22}$Na coincidence spectrum simulated with Geant4. The ROI of $\gamma^{\pm}$ // $\gamma^{\pm}$ coincidence events is highlighted in the insert at the top right of the figure. The intrinsic energy resolution of NaI(Tl) is not taken into account in the simulation and therefore lines are much sharper than in actual measurements.

### E. Data Acquisition Software

The Zynq PS of the Pixie-Net, running a full Ubuntu Linux operating system, acts as its own host PC equivalent. A number of C routines employing I/O functions provided by the *xillybus lite* FPGA core and C driver [29] are used to set up parameters for the FPGA pulse processing, read event data, build energy histograms, and store results to a local SD card or network drive. Results are also made available on webpages hosted by a local web server. The PTP timestamping in the DP83640 requires software control (the "PTP stack") to compute delays to the PTP master clock and adjust the local clock frequency accordingly. The open source software *LinuxPTP* [30] was used for that purpose, requiring minor reconfiguration of the Zynq Linux kernel to enable several PTP relevant kernel options. In addition, the open source software *mii-tool* [31] was adapted to communicate with the DP83640. This allows enabling of SyncE mode and configuring the DP83640 registers that control the outputs of PTP synchronized general purpose pins. For example, a pin can be configured to go logic high briefly whenever the internal 32 bit nanosecond counter rolls over to the next second, and so create a pulse-per-second (PPS) signal. A pin can also be configured to go logic high when both the 32 bit second and the 32 bit nanosecond counters match a user defined value, which can be used as a "data acquisition enabled" signal for the FPGA processing at a user defined time and date (to nanosecond precision). This is used to synchronously start data acquisition in different Pixie-Net modules. For a coordinated data acquisition, these local Linux programs on each Pixie-Net were executed remotely from a Linux PC via SSH calls from a shell script.

## III. RESULTS AND DISCUSSION

### A. Timing Performance

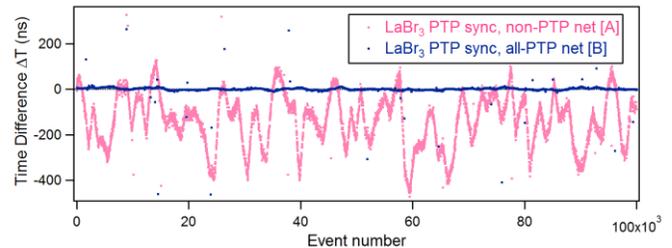

Fig. 8: Measured time difference ΔT for coincident pulses using PTP synchronization. Periodic adjustment in clock frequency by the PTP software cause increases or decreases in ΔT until the next adjustment.

Fig. 8 shows the measured ΔT distributions as a function of event number (corresponding to elapsed measurement time) for tests with PTP synchronization and two network configurations. True coincidences fall in a tight distribution; unrelated events close in time form a random background. The non-PTP network measurement clearly shows how ΔT drifts according to small clock frequency differences as the PTP software periodically adjusts the clock. This drift is of lower magnitude in the all-PTP network measurement, and not visible in SyncE measurements in any network configuration.

Histogramming hundreds of thousands of timing measurements, we obtain a distribution of ΔT around an average value (Fig. 9). The FWHM of a Gaussian fit to these distributions is the time resolution, our primary measure of performance for the synchronization methods and the network configuration (the switches).

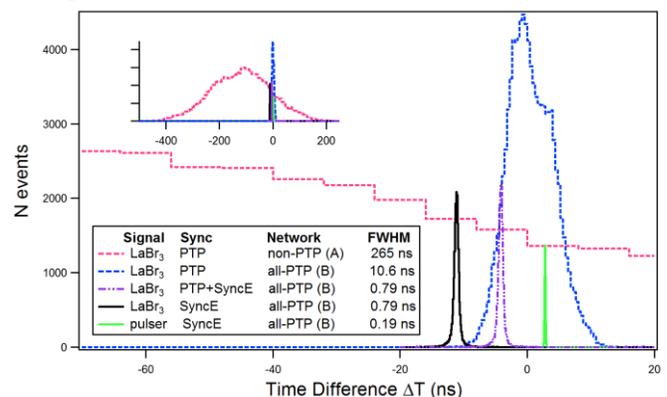

Fig. 9: Histograms of measured ΔT for various synchronization methods and network configurations. All-PTP networks have better time resolution that non-PTP networks, and SyncE synchronization is better again than PTP.

Fig.10 shows the time resolution for all measured methods and network configurations. The log scale bars indicate the measured time resolution; the y-axis labels indicate signal source and network type and switch; and the bar colors indicate the synchronization method.

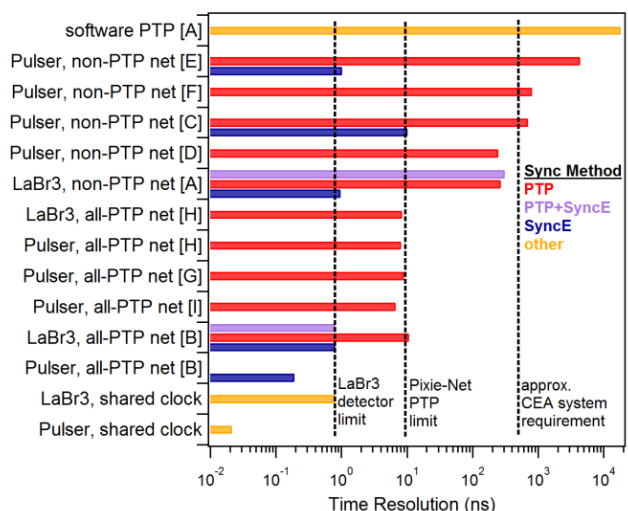

Fig. 10: Summary of timing resolutions for a variety of network switches, synchronization methods, and signal sources.

The reference measurement of a pure software PTP timestamping implementation (i.e. not using the DP38640) reaches ~18,000 ns time resolution and is clearly not acceptable for the current application.

Time resolutions for PTP hardware synchronization through non-PTP networks range from over 4000 ns to ~250 ns. Using all-PTP networks improves the timing resolution to 6-10 ns. We conclude that for PTP synchronization, the performance is dominated by the internal delays and latencies in the switches that vary significantly in different switch models. Though non-PTP switches are obviously not designed for highest timing performance, some reach acceptable levels of timing for this application. The PTP switches tested here perform much better, as expected, but are significantly more expensive.

Using SyncE synchronization improves the timing resolution to ~800 ps with the LaBr$_3$ detectors. (Note that the "all-PTP" networks are not using their PTP capabilities in SyncE only tests.) This value appears to be the detector limit in this particular setup, as also in a shared clock measurement no better than ~800 ps is reached with the LaBr$_3$ detectors, but ~20 ps is reached with a pulser. Pulser measurements with SyncE synchronization obtained timing resolutions in the range of 190-1000 ps for different switches. Combining SyncE and PTP gave mixed results; sometimes as good as SyncE only and sometimes worse than PTP only (when the PTP functions attempt to change the local clock frequency to compensate for measured time differences to the PTP master, even though the local clock is already synchronized in frequency by SyncE). This requires further study and likely can be improved by adjustments in the LinuxPTP settings or algorithms.

*B. Detector Coincidence Measurements*

At the CEA, a reference acquisition was made using an external shared clock in order to a) validate the simulations, and b) compare the resulting coincidence histogram (Fig. 11) with those obtained with two different switches. The time resolution (FWHM) of this reference configuration, measured by the time stamp differences of every coincident pulse pair, is ~20 ps. The time resolution for only NaI(Tl)//NaI(Tl) coincidence events is ~130 ns, likely due to the slower rise times and the pulse shape variations of detector signals (e.g. from light collection statistics).

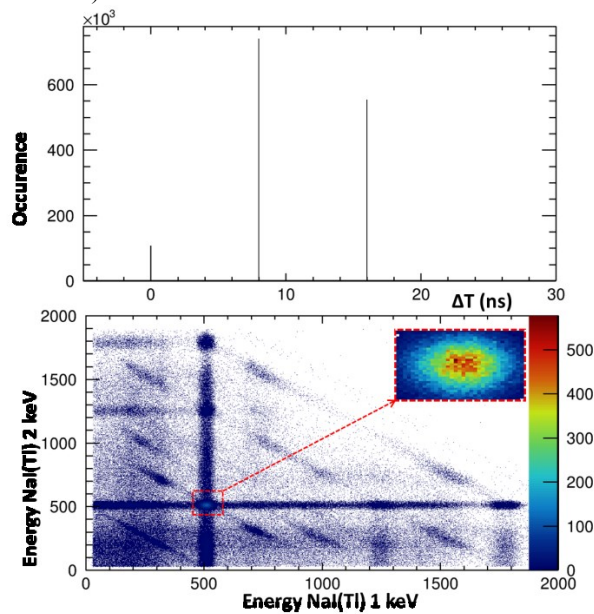

Fig. 11. Top: difference in pulse time stamps recorded by the two Pixie-Net modules. Bottom: $^{22}$Na coincidence spectrum obtained with a 0.2 μs coincidence window. the ROI of γ± // γ± coincidence events is highlighted in the insert at the top right of the figure. Results obtained with an external shared clock.

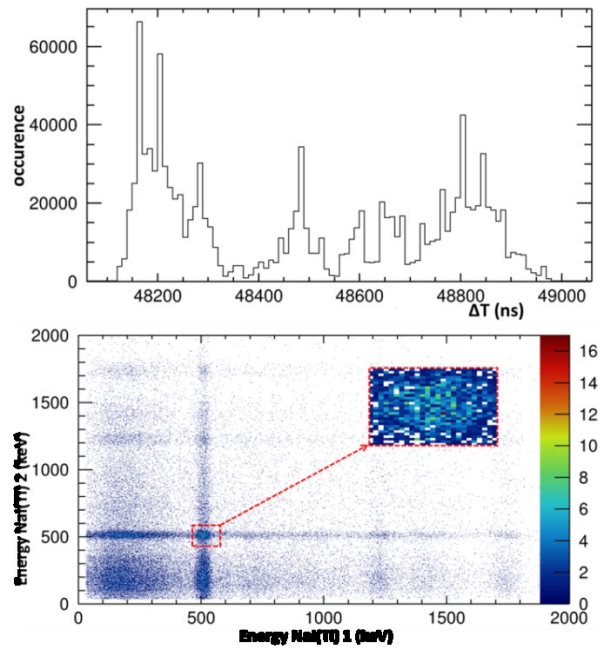

Fig. 12. Top: difference in pulse time stamps recorded by the two Pixie-Net modules. Bottom: $^{22}$Na coincidence spectrum obtained with a 49.5 μs coincidence window. The overwhelming majority of true coincidences are lost (compared to the simulated control spectrum); the ROI of γ± // γ± coincidence events is highlighted in the insert at the top right of the figure. Results obtained with a NetGear ProSafe GS108 switch (non PTP) [C].

The results from equivalent measurements with PTP synchronization and the NetGear ProSafe GS108 [C] switch demonstrates the timing precision's strong impact on the acquisition analysis: ΔT measured with this switch has a large offset (~ 50 µs) compared to the target coincidence window (0.8 – 1 µs), which can be corrected, but more significantly, the offset is not constant, see Fig. 12 (top). Reconstruction of coincidences is therefore rather complex and most events of interest are lost.

Indeed, as the bottom graph of Fig. 12 shows, no Compton scattering is visible, and less than 11 % of true coincidences could be recorded. Furthermore, fortuitous coincidences appear: two γ± emitted @ 511 keV from two different disintegrations are then counted in coincidence due to fluctuations in the switch offset).

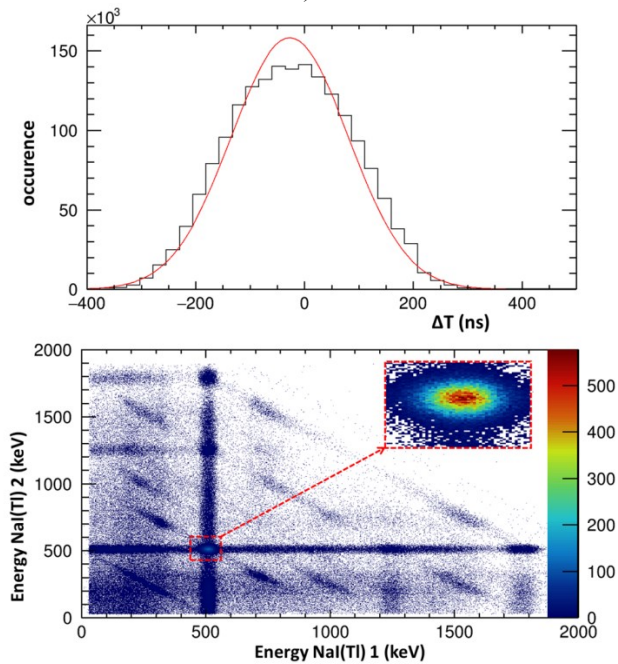

Fig. 13. Top : difference in pulse time stamps recorded by the two Pixie-Net modules. Bottom: $^{22}$Na coincidence spectrum obtained with a 0.32 µs coincidence window. All true coincidences could be reconstructed; the offset being constant, no fortuitous coincidence was added during the reconstruction of the events (with respect to the simulated control spectrum) the ROI of γ± // γ± coincidence events is highlighted in the insert at the top right of the figure. Results obtained with a Dell PowerConnect 2216 switch (non PTP) [A].

On the other hand, the results obtained with the Dell PowerConnect 2216 [A] switch (Fig. 13) show that ΔT measured with this switch has a small offset (<60 ns) and varies much less, with a 254 ns FWHM time accuracy. An optimized 320 ns time window was therefore applied to reconstruct the coincidences from the list mode files generated by each Pixie-Net module. Comparison of the total count in the γ± // γ± region of interest of the resulting spectrum with the total count in the same ROI of the control spectrum (obtained using shared clock) shows a difference of only 3.3% (Table II). Also, comparison of the total count in the γ± // γ± ROI (shared clock configuration) with the total count in the same ROI of the simulated spectrum (Fig. 6) presents a deviation of only 5 %. This gap can be explained by simulation simplifications, such as:

- Lack of knowledge of the dead zones and sensitive parts of the detectors;
- The incomplete understanding of the detector materials;
- The imperfect knowledge of the composition and geometry of the $^{22}$Na source used, as well as its 2% expanded relative uncertainty (k=2);
- A bias of random number generators;
- The environmental background is not considered in the simulation (which explains the experimental integral counts higher than the simulated counts reported in Table II.)
- Approximations of the physical theories considered by the simulation.

TABLE II
COINCIDENCE EVENTS RECORDED FOR A 7200 S ACQUISITION TIME WITH A $^{22}$NA SOURCE SANDWICHED BETWEEN THE TWO SQUARED NAI(TL) DETECTORS. THE TIME ACCURACIES (FWHM) MEASURED WITH EACH CONFIGURATION ARE REPORTED IN THE FIRST LINE.

| Counts | Geant4 | NetGear [C] PTP | Dell 2216 [A] PTP | Shared clock |
|---|---|---|---|---|
| Time resolution (ns) |  | ~690 | 254 | 129 |
| Integral | 313 540 | 58 540 | 332 420 | 334 960 |
| ROI 511 keV | 34 780 | 3 720 | 35 440 | 36 640 |

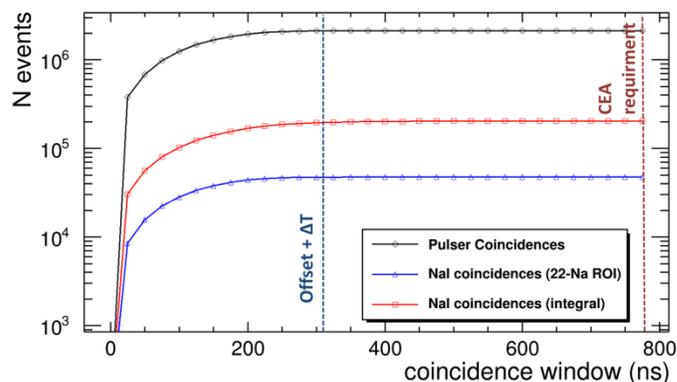

Fig. 14. Full coincidence counting for different coincidence window widths (post-analysis) in three regions: The black curve represents the pulse coincidences (channel 1 of the PN1/PN2 modules). The red curve represents all the coincidences recorded on channel 0 of PN1/PN2 modules. The blue curve represents the coincidences recorded in the γ± // γ± ROI on channel 0 of PN1/PN2 modules. Results obtained with the selected switch (Dell PowerConnect 2216 [A]).

These measurements show that good PTP synchronization is sufficient (even with a non-PTP switch), and allow to reliably reconstruct true coincidence events, without any loss, nor by adding fortuitous coincidences. The number of counts reported in Table II show a good agreement between the simulation, the shared clock (classic setup), and the PTP synchronization using a switch having a stable offset and an acceptable accuracy. Using this protocol can thus replace a shared external clock, which in our application effectively reduces the system's compactness and power consumption.

In addition, with a digital acquisition module like the Pixie-Net, the data are processed in post-analysis, which makes it possible to adjust the width of the coincidence window in order to study its impact on background noise rejection, see Fig. 14. With an analog acquisition chain, this window width is a parameter that must be set before starting the acquisition.

## IV. CONCLUSIONS AND OUTLOOK

As is to be expected, resolutions with an all-PTP network are better than with the non-PTP network, and better again with SyncE, by orders of magnitude. The hardware network time synchronization methods investigated here perform better than software timestamping, but not as good as shared clocks.

SyncE is a simple method as it requires only that the upstream network switch clocks its outputs from a common source, which can be expected to be a very common (since the easiest) architecture. However, as it provides only a synchronized clock frequency, not an absolute time reference, an additional signal is required to start and stop data acquisitions at the same time in different modules. In contrast, PTP is more complex and performs best with specialized switches, but allows data acquisition to a user defined date and time, with no extra cabling.

Overall, PTP and SyncE are promising network time synchronization techniques for nuclear physics applications. Though not reaching the time resolution from shared clocks, the measured resolutions of ~10ns (PTP) or < 1ns (SyncE), or even of ~250 ns (PTP with non-PTP switches), are sufficiently accurate for background coincidence rejection, as in the current application. The techniques studied are not suitable for time-of-flight measurements that desire picosecond timing, but similar methods could be, for example those developed in the White Rabbit project [32] that have been demonstrated to reach timing resolutions below 10 ps [33]. In initial tests equivalent to the pulser SyncE measurements, we reached time resolutions of ~160 ps with a White Rabbit demo kit [34], slightly better than the SyncE pulser results. This will be studied further in future work.

We note that besides reduced cabling complexity for setup with multiple data acquisition modules, a major advantage of the network timing techniques is that the network infrastructure can be chosen to match an application's timing requirements. This allows the use of lower cost non-PTP switches for less demanding applications while staying compatible with PTP switches in more demanding applications. However, as switch characteristics significantly affects performance, it is important to select and test a suitable model.

The comparison between the spectrum obtained with shared clock synchronization and the one obtained using the PTP synchronization implemented on the Pixie-Net (plus an adapted switch) highlights the efficiency of the protocol. For our needs, the switch timing resolution is sufficient and has no significant impact on the counting of true coincidences. This method is therefore validated for the application at CEA and has also allowed to optimize the compactness of the system.

In the current application, coincidences are detected through offline analysis of data recorded by multiple modules. It is however also possible to detect coincidence in (quasi) real time. With event data time stamped with date and time (if desired, related to global UTC by a GPS linked clock master) and local data being buffered in local memory, "software triggering" can replace hard wired trigger logic. For example, each module can send out minimal data packages (metadata) containing timestamps and other essential information. This metadata can be used by a central processor to make accept/reject decisions, which are communicated back to all modules. The response of the central processor has to be fast enough to process the combined *average* count rate in the system, but does not have to be immediate as events can be easily buffered in the modules' local memory and the software triggering works with the matched timestamps, not the time of arrival of the data. Initial tests indicate that transmissions time for metadata and decisions can be in the order or 250 µs (round trip) during which data can easily be buffered in local RAM (~1 GB) even at high count rates. The modules thus independently move their full data to long term storage or discard after receiving the accept/reject decision. The central processor may accumulate summary data for monitoring the acquisition. This kind of data acquisition, where much of the data is stored locally but is used globally for event selection, may be useful in large, distributed detector systems, and allows the use of local data acquisition modules rather than a large central rack.

REFERENCES

[1] S. Akkoyun et al, "AGATA—Advanced GAmma Tracking Array", NIM A 668 (2012), 26-58, doi.org/10.1016/j.nima.2011.11.081
[2] W. Hennig, H. Tan, M. Walby, P. Grudberg, A. Fallu-Labruyere, W. K. Warburton, C. Vaman, K. Starosta, D. Miller "Clock and Trigger Synchronization between Several Chassis of Digital Data Acquisition Modules", NIM B 261 (2007) 1000–1004
[3] A. Kimura, M. Koizumi, Y. Toh, J. Goto and M. Oshima, "Performance of a data acquisition system for a large germanium detector array", International Conference on Nuclear Data for Science and Technology 2007, https://doi.org/10.1051/ndata:07400
[4] F. Alessio, S. Baron, M. Barros Marin, J.P. Cachemiche, F. Hachon, R. Jacobsson and K. Wyllie, "Clock and timing distribution in the LHCb upgraded detector and readout system" Journal of Instrumentation, Volume 10, February 2015, https://doi.org/10.1088/1748-0221/10/02/C02033
[5] Bing He, Ping Cao, De-Liang Zhang, Qi Wang, Ya-Xi Zhang, Xin-Cheng Qi, Qi An, "Clock distribution for BaF2 readout electronics at CSNS-WNS", 2017 Chinese Phys. C 41 016104, https://doi.org/10.1088/1674-1137/41/1/016104
[6] W. Hennig, S. Asztalos, D. Breus, K. Sabourov, W.K. Warburton, "Development of 500 MHz Multi-Channel Readout Electronics for Fast Radiation Detectors", IEEE Trans. Nucl. Sci, Vol. 57, No. 4, August 2010, p. 2365-2370
[7] C. Hellesen, M. Skiba, G. Ericsson, E. Andersson Sundén, F. Binda, S. Conroy, J. Eriksson, M. Weiszflog, "Impact of digitization for timing and pulse shape analysis of scintillator detector signals", NIM A 720 (2013), 135-140
[8] W. K. Warburton, W. Hennig, "New Algorithms For Improved Digital Pulse Arrival Timing With Sub-GSps ADCs", IEEE Trans. Nucl. Sci Vol 64, No 12, Dec. 2017, p. 2938-2950; DOI: 10.1109/TNS.2017.2766074
[9] https://standards.ieee.org/findstds/standard/1588-2008.html
[10] www.ti.com/product/DP83640/technicaldocuments
[11] "Zynq-7000 AP SoC - Precision Timing with IEEE1588 v2 Protocol", www.xilinx.com
[12] www.xia.com/Pixie-Net.html
[13] T.W. Bowyer, S.R. Biegalski, M. Cooper, P.W. Eslinger, D. Haas, J.C. Hayes, H.S. Miley, D.J. Strom, V. Woods, "Elevated radioxenon detected remotely following the Fukushima nuclear accident", Journal of Environmental Radioactivity, Volume 102, Issue 7, 2011, Pages 681-687, https://doi.org/10.1016/j.jenvrad.2011.04.009.
[14] P. Achim , S. Generoso, M. Morin, P. Gross, G. Le Petit, C. Moulin, "Characterization of Xe-133 global atmospheric background: Implications for the International Monitoring System of the

Comprehensive Nuclear-Test-Ban Treaty", Journal of Geophysical Research: Atmospheres, Volume 121, 2016, Pages 4951–4966, https://doi.org/10.1002/2016JD024872.

[15] S. Generoso, P. Achim, M. Morin, P. Gross, G. Le Petit, C. Moulin, "Seasonal variability of Xe-133 global atmospheric background: Characterization and implications for the international monitoring system of the Comprehensive Nuclear-Test-Ban Treaty", Journal of Geophysical Research: Atmospheres, Volume 123, 2018, Pages 1865–1882, https://doi.org/10.1002/2017JD027765.

[16] https://www.nndc.bnl.gov/nudat2/decaysearchdirect.jsp?nuc=135XE&unc=nds

[17] McIntyre, J.I., Abel, K.H., Bowyer, T.W., Hayes, J.C., Heimbigner, T.R., Panisko, M.E., Reeder, P.L., Thompson, R.C., 2001. "Measurements of ambient radioxenon levels using the automated radioxenon sampler/analyzer (ARSA)". J. Radioanal. Nucl. Chem. 248 (3), 629-635. http://dx.doi.org/10.1023/A:1010672107749.

[18] Ringbom, A., Larson, T., Axelsson, A., Elmgren, K., Johansson, C., 2003. SAUNA a system for automatic sampling, processing, and analysis of radioactive xenon." NIM A 508 (3), 542-553. http://dx.doi.org/10.1016/s0168-9002(03)01657-7.

[19] W. Hennig, W. K. Warburton, A. Fallu-Labruyere, K. Sabourov, M. W. Cooper, J. I. McIntyre, A. Gleyzer, M. Bean, E. P. Korpach, K. Ungar, W. Zhang, P. Mekarski, "Development of a phoswich detector system for radioxenon monitoring", J. Radioanal. Nucl. Chem (2009) 282: 681. https://doi.org/10.1007/s10967-009-0181-9

[20] Farsoni, A.T., Alemayehu, B., Alhawsawi, A., Becker, E.M., 2013. "A phoswich detector with Compton suppression capability for radioxenon measurements". IEEE Trans. Nucl. Sci. 60, 456-464.

[21] Le Petit,G., Cagniant,A., Morelle,M., Gross,P., Achim,P., Douysset,G., Taffary,T., Moulin, C., 2013 "Innovative concept for a major breakthrough in atmospheric radioactive xenon detection for nuclear explosion monitoring", .J. Radioanal. Nucl. Chem. (2013) 298:1159-1169, https://doi.org/10.1007/s10967-013-2525-8

[22] K. Khrustalev, V. Yu Popov, Yu S. Popov, "Silicon PIN diode based electron-gamma coincidence detector system for Noble Gases monitoring", Applied Radiation and Isotopes, Volume 126, 2017, Pages 237-239, https://doi.org/10.1016/j.apradiso.2017.02.010.

[23] C. E. Cox, W. Hennig, A. C. Huber, W. K. Warburton, P. M. Grudberg, S. J. Asztalos, H. Tan, S. Biegalski , "A 24-element Silicon PIN diode detector for high resolution radioxenon measurements using simultaneous X-ray and electron spectroscopy," 2013 IEEE Nuclear Science Symposium and Medical Imaging Conference (2013 NSS/MIC), Seoul, 2013, pp. 1-7. doi: 10.1109/NSSMIC.2013.6829481

[24] http://www.nucleide.org/DDEP_WG/DDEPdata.htm

[25] http://scionix.nl

[26] S. Agostinelli et al., "Geant4—a simulation toolkit", "Nuclear Instruments and Methods in Physics Research Section A: Accelerators, Spectrometers, Detectors and Associated Equipment, Volume 503, 3, Pages 250-303, 2003, https://doi.org/10.1016/S0168-9002(03)01368-8

[27] J. Allison et al., "Recent developments in Geant4", Nuclear Instruments and Methods in Physics Research Section A: Accelerators, Spectrometers, Detectors and Associated Equipment, Volume 835, Pages 186-225, 2016, https://doi.org/10.1016/j.nima.2016.06.125

[28] http://zedboard.org/product/microzed

[29] http://xillybus.com/xillybus-lite

[30] http://linuxptp.sourceforge.net

[31] https://sourceforge.net/projects/net-tools/

[32] https://www.ohwr.org/projects/white-rabbit

[33] M. Lipiński, T. Włostowski, J. Serrano and P. Alvarez, "White Rabbit: a PTP application for robust sub-nanosecond synchronization," 2011 IEEE International Symposium on Precision Clock Synchronization for Measurement, Control and Communication, Munich, 2011, pp. 25-30. doi: 10.1109/ISPCS.2011.6070148

[34] http://sevensols.com/index.php/products/wr-len/

[A] Dell PowerConnect 2216, non-PTP
[B] back to back, PTP
[C] Netgear ProSAFE GS108, non-PTP
[D] Toplink TK 1005G, non-PTP
[E] Linksys EZXS55W , non-PTP
[F] Moxa EDS-405A-PTP, non-PTP (disabled)
[G] Moxa EDS-405A-PTP, PTP
[H] Oregano syn1588, PTP
[I] Artel Quarra 2800, PTP